\documentclass[a4paper,11pt]{article}
\pdfoutput=1 

\usepackage{jheppub} 

\usepackage{amssymb}
\usepackage{graphicx}
\usepackage{ulem}
\usepackage{bbold}
\usepackage{amsmath}
\newcommand*\diff{\mathop{}\!\mathrm{d}}
\newcommand{\isequiv}[1]{\underset{#1}{\simeq}}
\newcommand{\ftop}{t\bar{t}t\bar{t}}

\usepackage[dvipsnames]{xcolor} 

\usepackage{graphicx}

\usepackage{lineno} 

\usepackage{makecell}
\usepackage{subcaption}

\title{Invariant-mass threshold resummation for the production of four top quarks at the LHC}

\author[a]{Melissa van Beekveld,}
\author[b]{Anna Kulesza,}
\author[b]{Michele Lupattelli,}
\author[a]{Tommaso Saracco}

\affiliation[a]{Nikhef, Theory Group, Science Park 105, 1098 XG, Amsterdam, The Netherlands}
\affiliation[b]{Institute for Theoretical Physics, University of Münster,
Wilhelm-Klemm-Straße 9, \\ D-48149 Münster, Germany}

\emailAdd{mbeekvel@nikhef.nl}
\emailAdd{anna.kulesza@uni-muenster.de}
\emailAdd{michele.lupattelli@uni-muenster.de}
\emailAdd{tsaracco@nikhef.nl}

\abstract{Using invariant-mass threshold resummation, we compute the invariant-mass distribution and the total cross section for $\ftop$ production at the LHC with centre-of-mass energy of $13.6$~TeV at NLL$’$ accuracy. This accuracy includes next-to-leading logarithmic contributions together with relative $\mathcal{O}(\alpha_s)$ non-logarithmic terms present in the threshold limit.
We match the NLL$’$ results to NLO in QCD and to complete NLO with electroweak corrections.
We find that the inclusion of NLL$'$ soft-gluon corrections significantly reduces the size of the theoretical uncertainties and greatly improves the convergence of the predictions when considering various choices of renormalisation and factorisation scales.}

\dedicated{\rm MS-TP-25-17, Nikhef 2025-008}

\begin{document} 


\maketitle
\flushbottom

%
\section{Introduction}
\label{sec:introduction}
The production of four top quarks from a proton-proton collision at the Large Hadron Collider (LHC) is one of the rarest Standard Model (SM) processes that have been measured by the LHC experiments \cite{ATLAS:2018kxv, ATLAS:2020hpj, ATLAS:2021kqb, ATLAS:2023ajo, CMS:2019jsc, CMS:2019rvj, CMS:2023ftu}. The $pp\to\ftop$ process is of great interest for several reasons. First of all, it is sensitive to the strength and nature of the top-quark Yukawa coupling \cite{Cao:2016wib, Cao:2019ygh}. It can also be used to constrain the total width of the Higgs boson~\cite{ATLAS:2024mhs}. Furthermore, the total production cross section is predicted to be enhanced in several beyond-the-SM scenarios \cite{Darme:2018dvz,Toharia:2005gm,Craig:2016ygr,Dicus:1994bm,Farrar:1978xj,Beck:2015cga,Calvet:2012rk,Plehn:2008ae,Craig:2015jba, Abasov:2024mwk}, involving for example supersymmetric or multi-Higgs contributions. The production of the four top quarks can also constrain operators in the effective field theory framework \cite{Hartland:2019bjb,Ethier:2021bye,Aoude:2022deh,Zhang:2017mls,Aguilar-Saavedra:2018ksv,Banelli:2020iau,Darme:2021gtt}.

The fixed-order SM predictions with stable top quarks are known at next-to-leading order (NLO) accuracy. Pure QCD NLO corrections were calculated in \cite{Bevilacqua:2012em, Alwall:2014hca, Maltoni:2015ena}, whereas full QCD and electroweak (EW) NLO predictions were later obtained in \cite{Frederix:2017wme}. 
The first step towards including top decays was made in Ref.~\cite{Jezo:2021smh}, where the leading-order (LO) decay of the top quarks in the $1$-lepton + jets channel was considered and the result was matched to parton showers in the POWHEG~\cite{Nason:2004rx, Alioli:2010xd} framework.
Recently, theoretical predictions of $\ftop$ production in the four-lepton and three-lepton decay channels have been computed using the narrow-width approximation~\cite{Dimitrakopoulos:2024qib, Dimitrakopoulos:2024yjm}, meaning that the kinematics of the decay products and spin correlations are included at NLO accuracy. 

The cross section for the production of four top quarks has been recently measured by both the ATLAS \cite{ATLAS:2018kxv, ATLAS:2020hpj, ATLAS:2021kqb, ATLAS:2023ajo} and CMS \cite{CMS:2019jsc, CMS:2019rvj, CMS:2023ftu} collaborations.
The corresponding analyses are currently limited by statistical uncertainties. The latest
ATLAS measurement reads $22.5^{+4.7}_{-4.3}\text{(stat)}^{+4.6}_{-3.4}\text{(syst)}$ fb~\cite{ATLAS:2023ajo}, with a combined uncertainty of $29\%$, while the  CMS collaboration reports $17.7^{+3.7}_{-3.5}\text{(stat)}^{+2.3}_{-1.9}\text{(syst)}$ fb~\cite{CMS:2023ftu}, and is characterised by a combined uncertainty of $25\%$. These uncertainties are comparable in size with the uncertainties of the NLO theoretical predictions~\cite{Frederix:2017wme}.
Since the experimental uncertainties will be reduced as more data are collected with the high-luminosity phase of the LHC, bringing an order-of-magnitude larger luminosity, it is essential that the accuracy of the theoretical predictions improves as well.
Moreover, now that the $\ftop$ process has been observed, experimental collaborations will next target measurements of differential distributions, necessitating further theoretical work.

The theoretical description of the stable top-quark production can be further improved with the help of soft-gluon resummation. The use of resummation has been successful in obtaining more precise predictions for processes involving two heavy coloured final-state particles, such as $t\bar{t}$ \cite{Kidonakis:1997gm,Bonciani:1998vc,Kidonakis:2001nj, Czakon:2009zw,Beneke:2009rj,Ahrens:2010zv,Cacciari:2011hy,Beneke:2011mq,Czakon:2018nun} and $t\bar{t}H/Z/W^{\pm}$ production \cite{ Kulesza:2015vda, Kulesza:2016vnq, Kulesza:2017ukk, Kulesza:2018tqz, Kulesza:2020nfh, vanBeekveld:2020cat, Li:2014ula, Broggio:2015lya, Broggio:2016zgg, Broggio:2016lfj, Broggio:2017kzi, Broggio:2019ewu}. The production of $\ftop$ has been studied for the first time in the absolute-mass threshold formalism in Ref.~\cite{vanBeekveld:2022hty}.
 
 In the present work, we consider the resummation of logarithmic corrections that become large in the limit of the invariant mass of the four top quarks approaching the partonic centre-of-mass energy, instead of those becoming large at the production threshold, as considered in Ref.~\cite{vanBeekveld:2022hty}.
This approach allows also the calculation of all-order soft-gluon corrections to the invariant-mass distribution of the $\ftop$ final state.\footnote{Preliminary results at NLL accuracy for restricted ranges of the invariant mass were obtained in~\cite{Simon:2023xlo}.}
 
The rest of the paper is organised as follows. In Section~\ref{sec:theory} the general framework of invariant-mass threshold resummation is presented. In Section~\ref{sec:SAD} we focus on the treatment of the Coulomb limit in this kind of threshold resummation for processes with more than two final-state heavy coloured particles.
Numerical predictions for invariant-mass distributions and the total cross section are provided in Section~\ref{sec:results}.
Our results are summarised in Section~\ref{sec:sum}.
 
A comparison of the absolute-mass threshold resummation formalism with the invariant-mass one is provided in Appendix~\ref{sec:InvvsAbs}. In Appendix~\ref{sec:NbarvsN} we assess the differences between the $\bar{N}$- and $N$-resummation approaches.
Finally, in Appendix~\ref{sec:expanded}, we present approximate NLO predictions extracted from the resummation calculation and compare them to the exact NLO to assess the quality of the approximation.

%

\section{Theoretical framework} \label{sec:theory}
In threshold resummation one calculates the effects of soft-gluon emissions, which become important  in the limit of the partonic centre-of-mass energy $\sqrt{\hat{s}}$ approaching a certain threshold. 
In the presence of a multi-particle massive final state, one can design several threshold variables. A possible choice is the \textit{absolute-mass threshold}.
As the name suggests, this energy threshold is defined by the mass of the final-state system, $M \equiv 4m_t$ in our case. From this choice follows the definition of a dimensionless hadronic threshold variable $\rho_M=M^2/S$, and of the corresponding partonic threshold variable $\hat\rho_M=M^2/\hat{s}$.
Absolute-mass threshold resummation effectively includes soft-gluon corrections to the total cross section from the phase-space region where the final state is produced almost at rest.
It has been applied to the $pp \rightarrow \ftop$ process already in Ref.~\cite{vanBeekveld:2022hty}. 

Another possibility is the \textit{invariant-mass threshold}, where the threshold is given by the invariant mass of the final-state system $Q$. In this case, the hadronic and partonic threshold variables are $\rho_Q=Q^2/S$ and $\hat\rho_Q=Q^2/\hat{s}$, respectively. As a result, it becomes possible to study the effect of soft gluon corrections on the invariant-mass distribution of the final state. Considered for the total cross section, invariant-mass threshold resummation effectively includes soft-gluon corrections for all the invariant-mass configurations of the final state. We defer a more detailed discussion about the different threshold resummation approaches and their impact on the predictions to Appendix \ref{sec:InvvsAbs}.

In this work, invariant-mass threshold resummation is used (hence $\rho_Q\equiv\rho$ from now on, unless otherwise specified).
The resummed predictions are obtained at NLL$'$ accuracy, which includes the next-to-leading logarithmic (NLL) contributions together with relative $\mathcal{O}(\alpha_s)$ non-logarithmic contributions which do not vanish in the $\hat \rho_Q \to 1$ limit.

The hadronic differential cross section $\diff\sigma_{pp\to \ftop}/\diff Q$ can be written in terms of the partonic one as
\begin{equation} \label{eq:hadxsec}
    \frac{\diff\sigma_{pp\to \ftop}(\rho)}{\diff Q} = \sum_{i,j}\int \diff x_1 \diff x_2 \diff\hat\rho \,\delta(\hat\rho-\rho/x_1 x_2) f_{i}(x_1)f_{j}(x_2) \, \frac{\diff\sigma_{ij\to \ftop}(\hat\rho)}{{\diff Q}}\;,
\end{equation}
where $f_i$ are the parton distribution functions (PDFs), $x_i$ the momentum fractions and $\rho$ ($\hat\rho$) the hadronic (partonic) threshold variable. The sum over ${i,j}$ runs over the partonic channels, which are, at Born level, $q\bar{q}$ and $gg$.

At the matrix element level, the soft emissions factorise from the hard off-shell dynamics~\cite{Sterman:1986aj,Catani:1989ne,Contopanagos:1996nh,Kidonakis:1998nf}.
However, the factorisation of the phase space is only achieved in a conjugate space.
We therefore work in Mellin space, taking the Mellin transform of the cross section with respect to the hadronic threshold variable $\rho$, 
\begin{equation}
    \frac{\diff\tilde{\sigma}_{pp\to \ftop}(N)}{\diff Q}=\int_0^1 \diff\rho\, \rho^{N-1}\,\frac{\diff\sigma_{pp\to \ftop}(\rho)}{\diff Q}\;.
\end{equation}
In this way, the limit $N\to\infty$ isolates the threshold limit $\hat\rho \to 1$, turning off subleading $\mathcal{O}(1/N)$ corrections. Thus, plus-distributions involving $\ln(1-\hat\rho)$ are mapped into logarithms of $N$, or more specifically, logarithms of $\bar{N} = N e^{\gamma_E}$. Working with $\log \bar{N}$ terms instead of $\log N$ terms captures $\gamma_E$ terms that would otherwise be included only at higher logarithmic accuracy. This is the approach employed in this work, unless otherwise specified.
We defer a more detailed discussion of the difference due to the resummation of logarithms of $\bar{N}$ instead of $N$ to Appendix~\ref{sec:NbarvsN}.

The factorised expression of the resummed partonic cross section in $N$-space reads
\begin{equation} \label{eq:xsec_factorisation}
    \frac{\diff\tilde{\sigma}^{\text{res}}_{ij\to \ftop}(N)}{\diff Q} =  \;\text{Tr}\left[\bar{\textbf{S}}_{ij\to \ftop}(N+1)\; \textbf{H}_{ij\to \ftop}\right]\,\Delta_i(N+1)\Delta_j(N+1)\;.
\end{equation}
Hard-virtual and collinear endpoint corrections are contained in the hard function $\textbf{H}$, which is free from logarithmic enhancements. Collinear and soft-collinear radiation is taken into account in the two initial-state jet functions $\Delta_i$, whereas wide-angle soft-gluon emissions contribute to the soft function $\textbf{S}$. 
To avoid double counting the soft-collinear region both in the soft function and in the collinear functions, we subtract the eikonal jet functions $J^{\rm eik}_i$ from the soft function, $\bar{\textbf{S}}\equiv {\textbf{S}}/J^{\rm eik}_1 J^{\rm eik}_2$.
The soft and the hard functions are matrices in colour space. For $\ftop$ production, the colour space is $6$-dimensional for the $q\bar{q}$ channel and $14$-dimensional for the $gg$ channel~\cite{vanBeekveld:2022hty}. 

The hard function $\textbf{H}$ is finite in four dimensions and can be expanded in powers of the strong coupling $\alpha_s$,
\begin{equation} \label{eq:hardfuncexpansion}
    \textbf{H}_{ij\to \ftop} = \textbf{H}^{(0)}_{ij\to \ftop}+\frac{\alpha_s}{4\pi}\, \textbf{H}^{(1)}_{ij\to \ftop} + \mathcal{O}(\alpha_s^2)\;,
\end{equation}
where it is sufficient to take into account only the Born term $\textbf{H}^{(0)}$ at NLL accuracy. For NLL$'$ and higher accuracy, the $\textbf{H}^{(1)}$ term is needed. More specifically,
\begin{equation} \label{eq:H1eqV1plusC1}
    \textbf{H}^{(1)}_{ij\to \ftop} = \textbf{V}^{(1)}_{ij\to \ftop} + \textbf{C}^{(1)}_{ij\to \ftop}\;,
\end{equation}
where $\textbf{V}^{(1)}_{ij\to \ftop}$ are the one-loop virtual corrections and $\textbf{C}^{(1)}_{ij\to \ftop}$ accounts for relative order $\alpha_s$ (soft-) collinear $\log N$-independent contributions not captured by the NLL jet functions $\Delta_i$.

The collinear functions are computed analytically as \cite{Contopanagos:1996nh, Catani:1996yz}
\begin{equation} \label{eq:collinearfunctions}
    \Delta_i(N)=\exp \left\{\ \sum_{k=1}^\infty \alpha_s^{k-2} g_k(\lambda)
    \right\},
\end{equation}
where $\lambda=\alpha_s b_0 \log \bar{N}$, and at NLL (and NLL$'$) accuracy the functions up to $g_2(\lambda)$ are needed (see e.g.
\cite{Catani:2003zt, vanBeekveld:2019cks}).

The soft function is obtained by solving the corresponding renormalisation-group equation, and can be written as \cite{Contopanagos:1996nh, Kidonakis:1997gm, Kidonakis:1998nf, Kidonakis:1998bk}
\begin{equation}
\label{eq:soft_f_ev}
    {\textbf{S}}_{ij\to \ftop} = \bar{\textbf{U}}_{ij\to \ftop} \,\tilde{{\textbf{S}}}_{ij\to \ftop} \,\textbf{U}_{ij\to \ftop} \;,
\end{equation}
where $\tilde{{\textbf{S}}}$ is the boundary condition evaluated at the soft scale $\mu_R=Q/\bar{N}$ and $\textbf{U}$ is the soft function evolution matrix.
The boundary condition $\tilde{{\textbf{S}}}$ can be expanded perturbatively in powers of $\alpha_s$
\begin{equation}
\tilde{{\textbf{S}}}_{ij\to \ftop} = \tilde{{\textbf{S}}}^{(0)}_{ij\to \ftop} + \frac{\alpha_s}{4\pi}\, \tilde{{\textbf{S}}}^{(1)}_{ij\to \ftop} + \mathcal{O}(\alpha_s^2).
\end{equation}
The lowest-order contribution is $\tilde{{\textbf{S}}}^{(0)}$, whose matrix elements are given by $\text{Tr}(c_I c_J^\dagger)$, where $c_I$ and $c_J$ are colour basis vectors.
The one-loop correction $\tilde{{\textbf{S}}}^{(1)}$, needed at NLL$'$, needs to be calculated analytically from one-loop eikonal emissions \cite{Korchemsky:1993uz, Catani:1996vz, Ahrens:2010zv, Lyubovitskij:2021ges}.
The soft evolution matrix can be written as a path-ordered exponential \cite{Kidonakis:1998nf}
\begin{equation} \label{eq:softevolutionmatrix}
    \textbf{U}_{ij\to t \bar t t \bar t}=\mathcal{P}\exp\left[\frac12 \int_{\mu_R^2}^{Q^2/\bar{N}^2} \frac{\diff \mu^2}{\mu^2} \,{\bf\Gamma}_{ij\to t \bar t t \bar t}(\mu^2,\alpha_s(\mu^2)) \right],
\end{equation}
where ${\bf\Gamma}_{ij\to t \bar t t \bar t}$ is the soft anomalous dimension, which is a finite function governing the soft behaviour of the process.
Both at NLL and NLL$'$ only the first order coefficient in the expansion of ${\bf\Gamma}$, ${\bf\Gamma}=  \frac{\alpha_s}{4\pi}\, {\bf\Gamma}^{(1)} + \dots$, is needed. This coefficient can be calculated perturbatively from the UV poles of one-loop eikonal contributions.
Eq.~\eqref{eq:softevolutionmatrix} can be reduced to an ordinary exponential if the soft anomalous dimension is diagonal.
This can be achieved by finding an appropriate diagonalisation matrix on a phase-space point basis.
We comment on the technical aspects of the calculation of the soft anomalous dimension in Sec.~\ref{sec:SAD}.

Once all the ingredients of Eq. \eqref{eq:xsec_factorisation} have been computed, we need to go back to momentum space, which is done via an inverse Mellin transform
\begin{equation}
\label{eq:had_res_diff_xs}
    \frac{\diff \sigma^{\text{res}}_{pp\to \ftop}(\rho)}{\diff Q}=\sum_{i,j}\int_{\mathcal{C}}\frac{\diff N}{2\pi i}\, \rho^{-N} f_i(N+1)f_j(N+1) \,\frac{\diff \tilde{\sigma}^{\text{res}}_{ij\to \ftop}(N)}{\diff Q}\;.
\end{equation}
To perform the inverse Mellin transform we make use of  the Minimal Prescription (MP) method \cite{Catani:1996yz}. The integral is calculated numerically, parametrising the contour $\mathcal{C}$ as $N=C_{MP}+y e^{i\phi_{MP}}$ with $y\in[0,\infty)$. 

To further improve the accuracy of the result, we match the resummed prediction to the exact NLO result. To avoid double counting, we subtract the resummed result expanded up to NLO. The general formula for additive matching is given by
\begin{equation} \label{eq:matching}
    \diff \sigma^{\text{f.o.}+\text{res}}=\diff\sigma^{\text{f.o.}} + \left[\diff\sigma^{\text{res}}-\diff\sigma^{\text{res}}|_{\mathcal{O}(\alpha_s^n)}\right],
\end{equation}
where ``f.o." denotes the fixed-order (LO, NLO, ...) contribution, and ``res" indicates the resummation accuracy (LL, NLL, NLL$'$, ...). The term $\sigma^{\text{res}}|_{\mathcal{O}(\alpha_s^n)}$ is the expansion of the resummed cross section up to $\alpha_s^n$, where $n$ is the same power as in the fixed-order term ($n=4$ at LO and $n=5$ at NLO).

%

\section{Treatment of the Coulomb limit}

\label{sec:SAD}

As discussed in the previous section, emissions of soft gluons close to the absolute threshold limit, i.e.\ when all (heavy, coloured) particles in the final state are produced almost at rest, lead to large logarithmic corrections. In this limit another type of corrections, known as Coulomb corrections, becomes relevant too. The Coulomb contributions originate from virtual gluon exchanges between slowly moving particles, and are therefore to be also expected when only a subset of final-state particles has low velocity, as it can be the case for production of a system with an invariant mass $Q>M$ . It has been shown for the case of a pair production of heavy coloured particles that Coulomb corrections factorise from the hard and soft functions \cite{Beneke:2010da} and can be resummed~\cite{Fadin:1990wx,Catani:1996dj,Beneke:2010da,Kulesza:2009kq}. Resummation of Coulomb corrections for the four-top production process is, however, beyond the scope of the presented work. Here we include Coulomb corrections at one-loop level as part of the NLO input.\footnote{Note that, as at NLL$'$ the hard function involves the one-loop virtual corrections, we will include cross terms consisting of the NLO Coulomb terms with soft logarithms.}

Nevertheless, there is another step in the calculation where we are sensitive to the Coulomb effects, namely in the computation of the one-loop soft anomalous dimension. Coulomb singularities arise in ${\bf \Gamma}^{(i)}$  when (a subset of)  heavy coloured particles are produced with zero velocity. To see it explicitly, let us consider the one-loop soft anomalous dimension for a generic $2\to n$ process, with a massless initial state, either $q \bar{q}$ or $gg$, and with $n$ massive coloured final-state particles~\cite{Becher:2009kw}
\begin{align} \label{eq:SAD1loop}
\begin{split}
    {\bf\Gamma}^{(1)}_{2 \to n}&=\,
\textbf{T}_{1}\cdot\textbf{T}_{2}\,\gamma^{(0)}_{\text{cusp}}\log\left(\frac{\mu^2}{-s_{12}}\right) + \sum_{i=1}^2 \gamma_i^{(0)}\,\mathbf{1}+\sum_{I=3}^n \gamma_I^{(0)}\,\mathbf{1} \\
    &\quad- \frac12 \sum_{\substack{I,J=3}}^n \textbf{T}_{I}\cdot\textbf{T}_{J}\, \gamma^{(0)}_{\text{cusp}}(\beta_{IJ}) + \sum_{i=1}^2\sum_{J=3}^n \textbf{T}_{i}\cdot\textbf{T}_{J}\,\gamma^{(0)}_{\text{cusp}} \log\left(\frac{m_J\mu}{-s_{iJ}}\right), 
\end{split}
\end{align}
where the lowercase indices $i\in\{1,2\}$ indicate massless particles, the uppercase indices $I,J\in\{3,4,...,n\}$ refer to massive ones, and $m_J$ is the mass of the $J$-th quark in the final state. The one-loop cusp anomalous dimension coefficient is $\gamma^{(0)}_{\text{cusp}}=4$ and the one-loop anomalous dimension coefficients for massless quarks and gluons, $\gamma_i^{(0)}$, are given by $\gamma_q^{(0)}=-3\,C_F$ and $\gamma_g^{(0)}=-\frac{11}{3}\,C_A+\frac23 n_f$, respectively. The one loop coefficient for heavy quarks reads $\gamma_I^{(0)}=\gamma_Q^{(0)}=-2\,C_F$.
The kinematic dependence on the four-momenta enters via the Mandelstam variables $s_{12}=2p_1\cdot p_2+i0$ and $s_{iJ}=-2\,p_i\cdot p_J+i0$, and via the cusp angle $\beta_{IJ}$.
The one-loop massive cusp anomalous dimension reads
\begin{equation} \label{eq:cuspmass}
    \gamma^{(0)}_{\text{cusp}}(\beta_{IJ})=\gamma^{(0)}_{\text{cusp}}\,\beta_{IJ}\coth\beta_{IJ} \;,
\end{equation}
where 
\begin{equation}
	 \label{eq:coshbeta}
	\cosh \beta_{IJ}\equiv-\frac{p_I\cdot p_J}{m_I\,m_J}-\mathrm{i} 0\;, \quad  \beta_{IJ}=b_{IJ}-i\pi\,.
\end{equation} 
The $-\mathrm{i}0$ regulator is chosen in order to stay below the branch cut when we solve Eq.~\eqref{eq:coshbeta} for $\beta_{IJ}$, with $b_{IJ}$ real and positive.%
\footnote{The choice to stay below or above the branch cut is arbitrary and although it leads to opposite signs for the $\beta_{IJ}$ angle, it does not impact our results.}
In the limit of small $b_{IJ}$ we obtain
\begin{equation} \label{eq:cuspmassexpansion}
	\beta_{IJ}\coth\beta_{IJ}\isequiv{b_{IJ}\ll 1} \frac{(-\mathrm{i}\,\pi)}{b_{IJ}}+1 +\mathcal{O}(b_{IJ})\;,
\end{equation}
where the first term generates a Coulomb singularity in the limit $b_{IJ}\to 0$.

\begin{figure}[t]
    \centering
    \begin{equation*}
    \begin{gathered}
    \includegraphics[height=4cm]{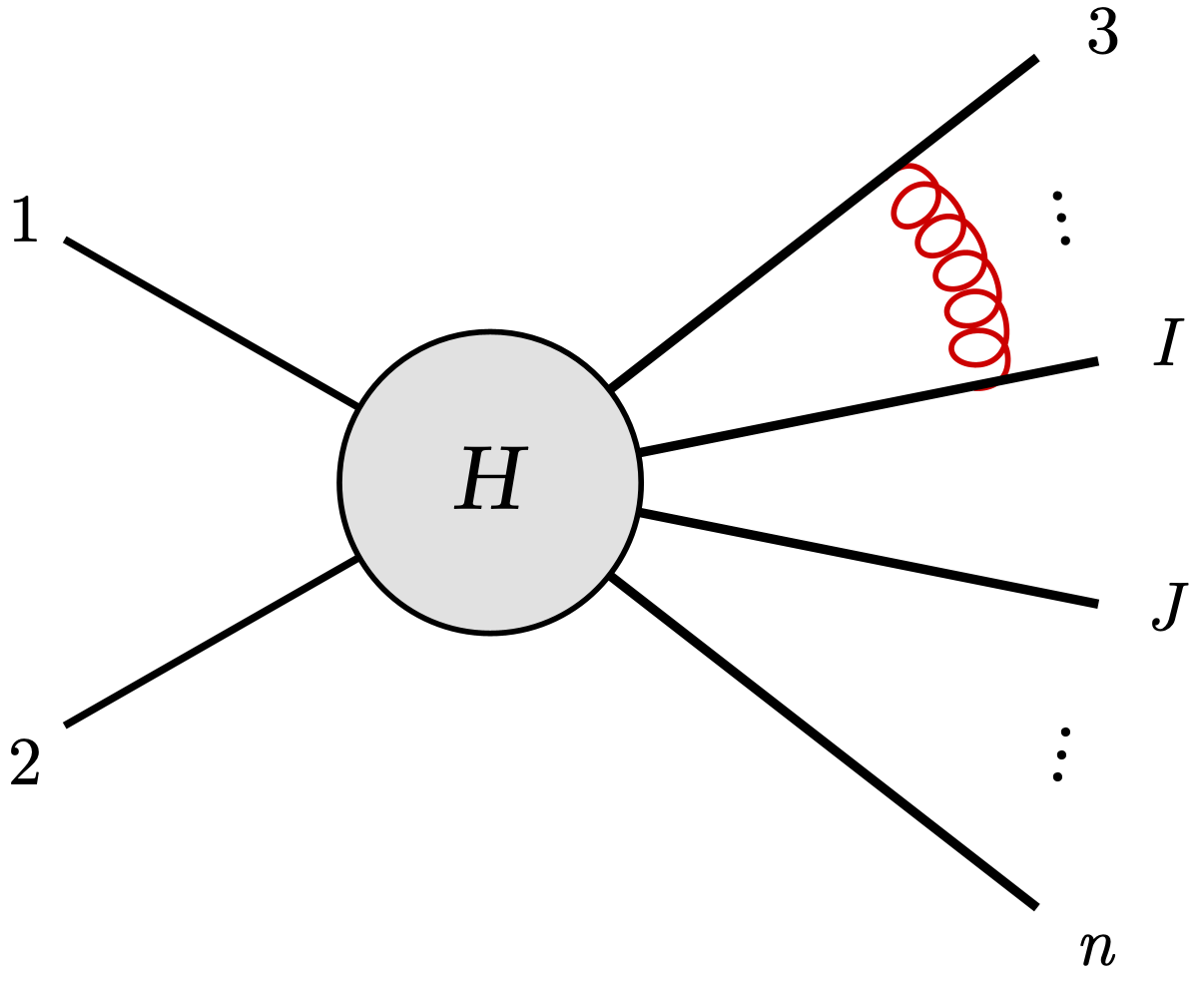}
    \end{gathered} \qquad \quad \text{\LARGE{$\longrightarrow$}} \qquad \quad  
    \begin{gathered}
    \includegraphics[height=4cm]{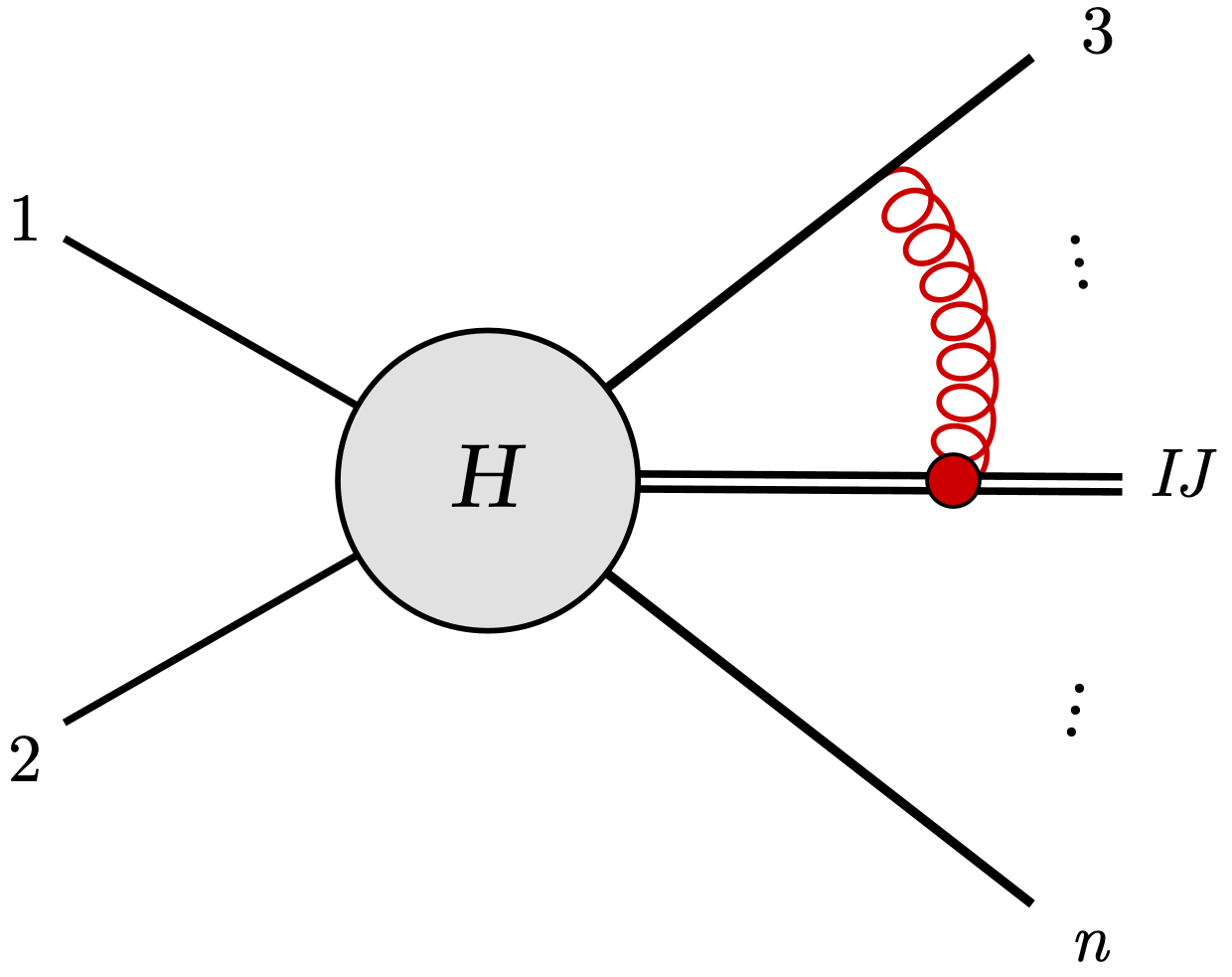}
    \end{gathered}
    \end{equation*}
    \caption{\it Pictorial representation of soft-gluon exchanges when two massive particles are emitted at rest. The incoming lines $1,2$ stand for either $q\bar{q}$ or $gg$ channel. The grey area denotes the hard function and the red gluon illustrates a virtual soft emission.
    In the limit where two final-state coloured particles $I$ and $J$ are emitted at rest, soft gluons effectively resolve them as a single particle in the tensor product representation.}
\label{fig:partialthresholdlimit}
\end{figure}

We note that Coulomb singularities do not pose a problem in absolute mass threshold resummation. One can always find  a colour basis in which the soft anomalous dimension matrix is diagonal, in which case only its real part contributes to the soft function, see Eqs.~\eqref{eq:soft_f_ev} and~\eqref{eq:softevolutionmatrix}. Moreover, the real parts of the eigenvalues correspond to the colour charges given by the Casimir invariants of the irreducible representations into which the tensor product representation of the final state decomposes, as has been explicitly shown for e.g.\ the top quark-pair~\cite{Bonciani:1998vc, Kidonakis:1997gm}, $\ftop$~\cite{vanBeekveld:2022hty} or supersymmetric particle~\cite{Kulesza:2009kq, Beenakker:2010nq} production. This, in turn, is in agreement with the physical picture in which soft gluons cannot resolve the final-state particles produced at rest and can only couple to the total colour charge of the final state~\cite{Bonciani:1998vc, Beneke:2009rj, Czakon:2009zw}. The same physical behaviour is to be expected when only one pair $IJ$, out of multiple number of heavy quarks in the final state, has almost no velocity, see Fig.~\ref{fig:partialthresholdlimit}. Given $I$ and $J$  in the representations $R$ and $R'$, soft gluons will only perceive an effective single particle in the $R\otimes R'$ representation, with generator $\textbf{T}_{IJ}=\textbf{T}_{I}+\textbf{T}_{J}$. Correspondingly, the soft anomalous dimension in this limit can be obtained from Eq.~\eqref{eq:SAD1loop} by lowering the upper limits in the sums over $I,J$ from $n$ to $n-1$ and introducing an anomalous dimension for the effective $IJ$ particle   ${\boldsymbol\gamma}^{(0)}_{IJ}=   \textbf{T}_{IJ}^2\, \gamma^{(0)}_{Q} /C_F$ using Casimir scaling. This can be also seen explicitly by considering the problematic $  \gamma^{(0)}_{\text{cusp}}(\beta_{IJ})$ in the small $b_{IJ}$ limit
\begin{align} \label{eq:SADthrpair}
     -\textbf{T}_{I}\cdot\textbf{T}_{J}\,\gamma^{(0)}_{\text{cusp}}(\beta_{IJ})  \isequiv{b_{IJ}\ll 1}& -\textbf{T}_{I}\cdot\textbf{T}_{J}\,\gamma^{(0)}_{\text{cusp}}\frac{(-\mathrm{i}\,\pi)}{b_{IJ}}  -\textbf{T}_{I}\cdot\textbf{T}_{J}\,\gamma^{(0)}_{\text{cusp}} + \mathcal{O}(b_{IJ}) \\
     =\hspace{0.75em}&  -\textbf{T}_{I}\cdot\textbf{T}_{J}\,\gamma^{(0)}_{\text{cusp}}\frac{(-\mathrm{i}\,\pi)}{b_{IJ}}  -2\gamma^{(0)}_{Q}\,\mathbf{1}+{\boldsymbol\gamma}^{(0)}_{IJ} + \mathcal{O}(b_{IJ})\;, \nonumber
\end{align}
where we used $\gamma^{(0)}_{\text{cusp}}=-2\gamma^{(0)}_{Q}/C_F$ and the fact that $\textbf{T}_{IJ}^2=\textbf{T}_{I}^2+\textbf{T}_{J}^2+2\,\textbf{T}_{I}\cdot\textbf{T}_{J}$ by colour conservation. 
Inserting the finite terms of the second line of Eq.~\eqref{eq:SADthrpair} in Eq.~\eqref{eq:SAD1loop}, we obtain the cancellation of the two slowly-moving-particle contributions ($I$,$J$), which are replaced by the anomalous dimension for the new effective particle $IJ$.\footnote{We remind the reader that the Coulomb-enhanced contributions are already accounted for in the hard function $\textbf{H}$.}
To take this limit in our numerical calculation, we use instead the finite part of the first line of Eq.~\eqref{eq:SADthrpair} and introduce a small parameter $\delta$ as a boundary value of $b_{IJ}$ below which this replacement takes place, i.e.
\begin{equation} \label{eq:eq:cuspmassreg}
    \gamma^{(0)}_{\text{cusp}}(\beta_{IJ})=\gamma^{(0)}_{\text{cusp}}\times \left\{
     \begin{array}{@{}l@{\thinspace}l}
       \beta_{IJ}\coth\beta_{IJ}   &\quad \text{for } b_{IJ}\geq \delta\,, \\
        1&\quad \text{for } b_{IJ} < \delta \;.
     \end{array}
     \right.
\end{equation}
This treatment naturally extends when larger subsets of final-state particles are emitted at rest.
 
%

\section{Results}
\label{sec:results}
In this section we present our results for the invariant-mass threshold resummation of the $\ftop$ process at NLO+NLL$'$ level, with and without inclusion of EW corrections. 
After introducing our setup in Section~\ref{sec:setup}, we show our results for the invariant-mass distribution for the $\ftop$ final state in Section~\ref{sec:inv-mass}, and the total cross section in Section~\ref{sec:total}.
Tests of the quality of the logarithmic approximation are detailed in Appendix~\ref{sec:expanded}.

\subsection{Input parameters and computational setup}
\label{sec:setup}
We present results for proton-proton collisions at the LHC with a centre-of-mass energy of 13.6~TeV, unless otherwise specified. 
The top-quark pole mass is set to 172.5~GeV.
For the electroweak parameters, we use the $G_\mu$ scheme, with $G_F = 1.166390\times 10^{-5}~{\rm GeV}^{-2}$, $m_W = 80.419~{\rm GeV}$ and $m_Z = 91.188~{\rm GeV}$. 
We use the five-flavour scheme and the LUXqed\_plus\_PDF4LHC15\_nnlo\_100 PDF set \cite{Manohar:2016nzj,Manohar:2017eqh}, and extract $\alpha_s(\mu)$ from the LHAPDF interface~\cite{Buckley:2014ana}.
This PDF set is based on the PDF4LHC15 PDF set \cite{Butterworth:2015oua,NNPDF:2014otw,Harland-Lang:2014zoa,Dulat:2015mca} and includes the photon content of the proton, necessary when computing the EW corrections.

By default, the renormalisation ($\mu_R$) and the factorisation ($\mu_F$) scales are set to a common value $\mu_0$.
An estimate of the scale uncertainty is performed by means of the \textit{7-point scale variation}, i.e.\ varying the scales in the range
\begin{equation}
	\label{eq:7-pt-var}
	\biggl( \frac{\mu_R}{\mu_0}, \frac{\mu_F}{\mu_0} \biggr) \in \bigl\{ (0.5,0.5), (0.5,1), (1,0.5), (1,1), (1,2), (2,1), (2,2) \bigr\}.
\end{equation}
The resummed predictions are calculated with an in-house code that makes use of colour-decomposed tree-level and one-loop amplitudes provided in a custom version of \textsf{OpenLoops} \cite{Ossola:2007ax,vanHameren:2009dr,vanHameren:2010cp,Cascioli:2011va,Denner:2016kdg,Buccioni:2017yxi,Buccioni:2019sur}. 
We have confirmed that the dependence of the results on the choice of the MP contour (Eq.~\eqref{eq:had_res_diff_xs}) for $C_{\rm MP}\in \{1.9,2.0,2.1,2.2\}$ and $\phi_{\rm MP}\in \{0.75,0.8,0.85\}\pi$, and on the choice of $\delta \in \{0.1,0.01,0.001\}$ (Eq.~\eqref{eq:eq:cuspmassreg}) is negligible numerically. 
For the predictions shown in the rest of this section we take $C_{\rm MP} = 2.1$, $\phi_{\rm MP} = 0.75\pi$ and $\delta = 0.1$.

The resummed predictions are matched with fixed-order results obtained using \textsf{MG5\_aMC@NLO} v3.5.5 \cite{Alwall:2014hca, Frederix:2018nkq}.
For our fixed-order results we consider pure QCD contributions of $\mathcal{O}(\alpha_s^4)$ and $\mathcal{O}(\alpha_s^5)$, and EW corrections~\cite{Frederix:2017wme} of $\mathcal{O}(\alpha_s^3\alpha)$, $\mathcal{O}(\alpha_s^2\alpha^2)$,
$\mathcal{O}(\alpha_s^4\alpha)$ and $\mathcal{O}(\alpha_s^3\alpha^2)$.
We label this accuracy as ``NLO".

\subsection{Invariant-mass distribution}
\label{sec:inv-mass}

\begin{figure}[t]
    \centering
    \includegraphics[width=0.45\textwidth]{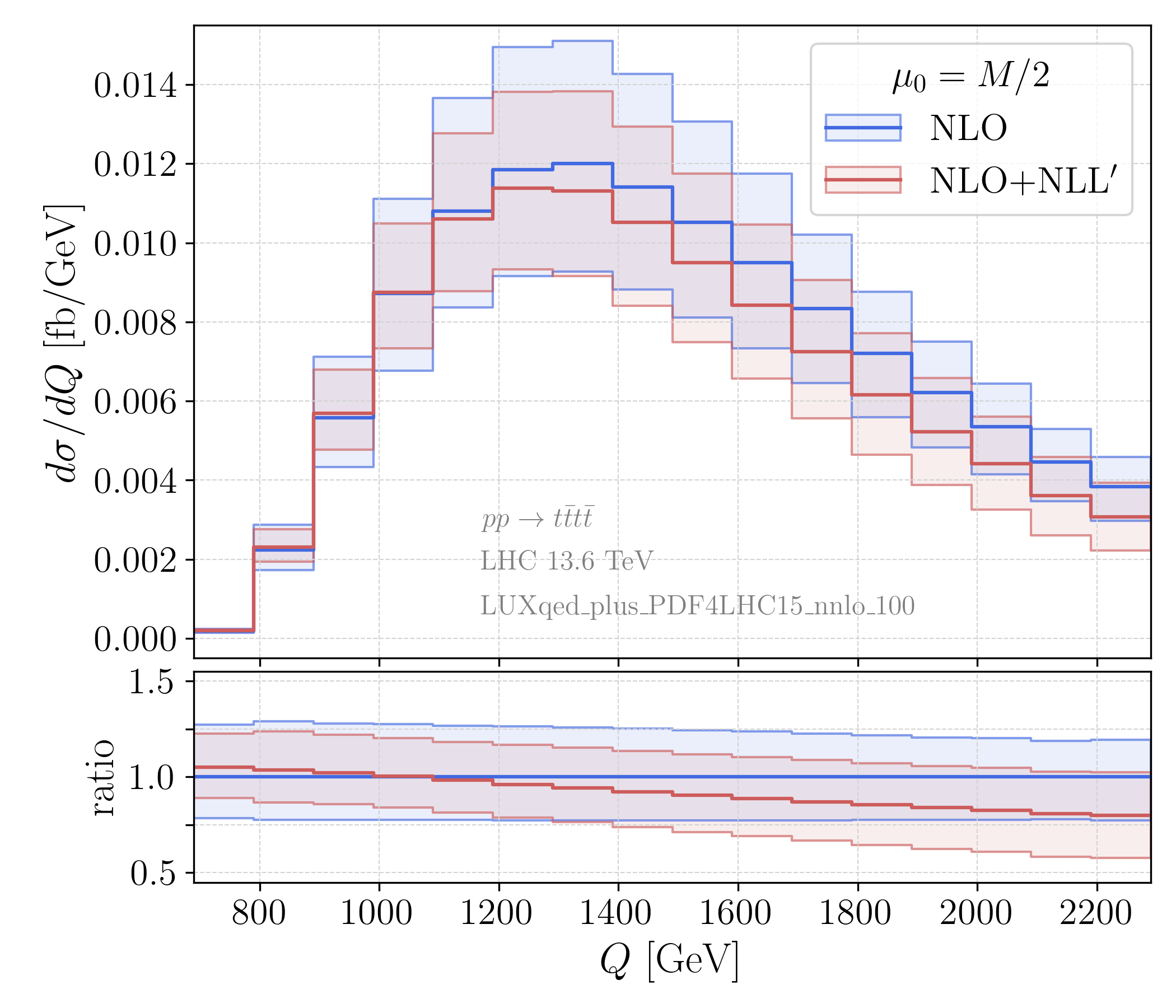}
    \includegraphics[width=0.45\textwidth]{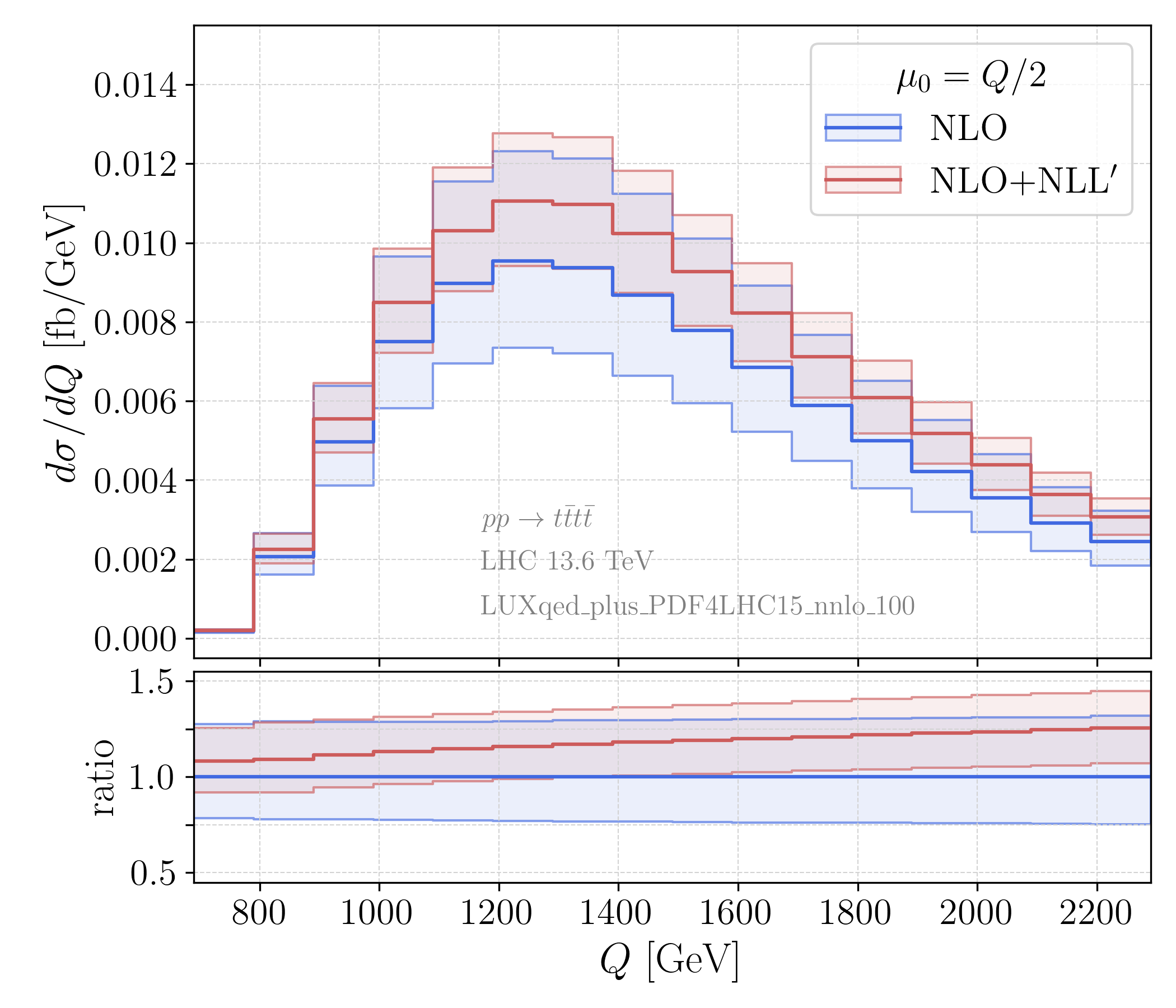}
    \includegraphics[width=0.45\textwidth]{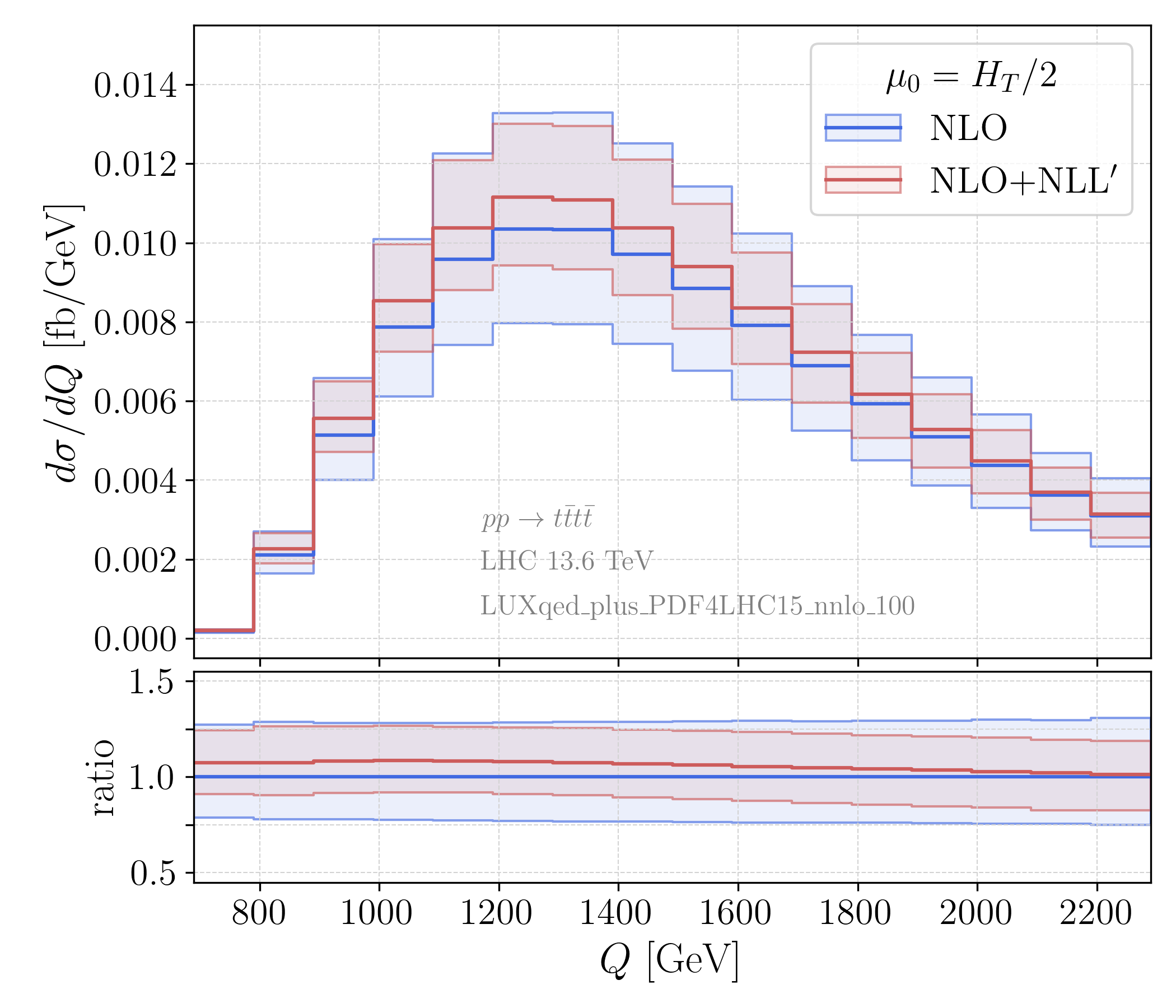}
    \caption{\it Invariant-mass distribution of the $t\bar{t}t\bar{t}$ system at NLO and at NLO+NLL$'$ accuracy.
    Results are presented for three different scales.
    The upper panels show the absolute distributions, the lower panels the distributions normalised bin-by-bin to the NLO distribution.
    }
    \label{fig:DiffDistrHT2}
\end{figure}

We start our presentation of the results by considering the invariant-mass distribution of the $\ftop$ final state. 
Fig.~\ref{fig:DiffDistrHT2} compares the NLO (blue) and NLO+NLL$'$ (red) distributions for three different scale choices:
$\mu_0 = M/2$ with $M=4m_t$ the total mass of the final state, $\mu_0 = Q/2$ with $Q$ the invariant mass of the final state, $\mu_0 = H_T/2$ with $H_T= \sum_{i=1}^4\sqrt{m_t^2 + p_{T,i}^2}$,
where the sum runs over the four top quarks, and $p_{T,i}$ is their transverse momentum in the laboratory frame. 
For each plot, the top panel displays the distributions themselves, whereas the lower panel shows the ratios to the NLO distribution. 
The scale uncertainty is indicated by the shaded coloured band. 

%

Almost all NLO and NLO+NLL$'$ distributions peak around the $Q = 1240$~GeV bin, except for the NLO distribution with $\mu_0 = M/2$, which peaks at the slightly higher bin value of $Q=1340$~GeV.
One may observe that the NLL$'$ corrections induce a modification of the invariant-mass distribution shape for all scale choices.
For the choice $\mu_0 = Q/2$ ($\mu_0 = M/2$) the change in shape is substantial, with corrections varying in the range $[8\%, 25\%]$ ($[5\%,-20\%]$) in the phase-space region studied.
For $\mu_0 = M/2$, they start off positive and then change sign before reaching the bulk of the distribution, becoming increasingly negative thereafter, whereas for $\mu_0 = Q/2$ the corrections are increasingly positive.
A notably smaller shape modification is observed for $\mu_0 = H_T/2$, with NLO+NLL$'$ corrections ranging from $1\%$ to $8\%$.
For all three scale choices we observe that the scale uncertainty is substantially reduced for the NLO+NLL$'$ predictions.

\begin{figure}[t]
    \centering
    \includegraphics[width=0.45\textwidth]{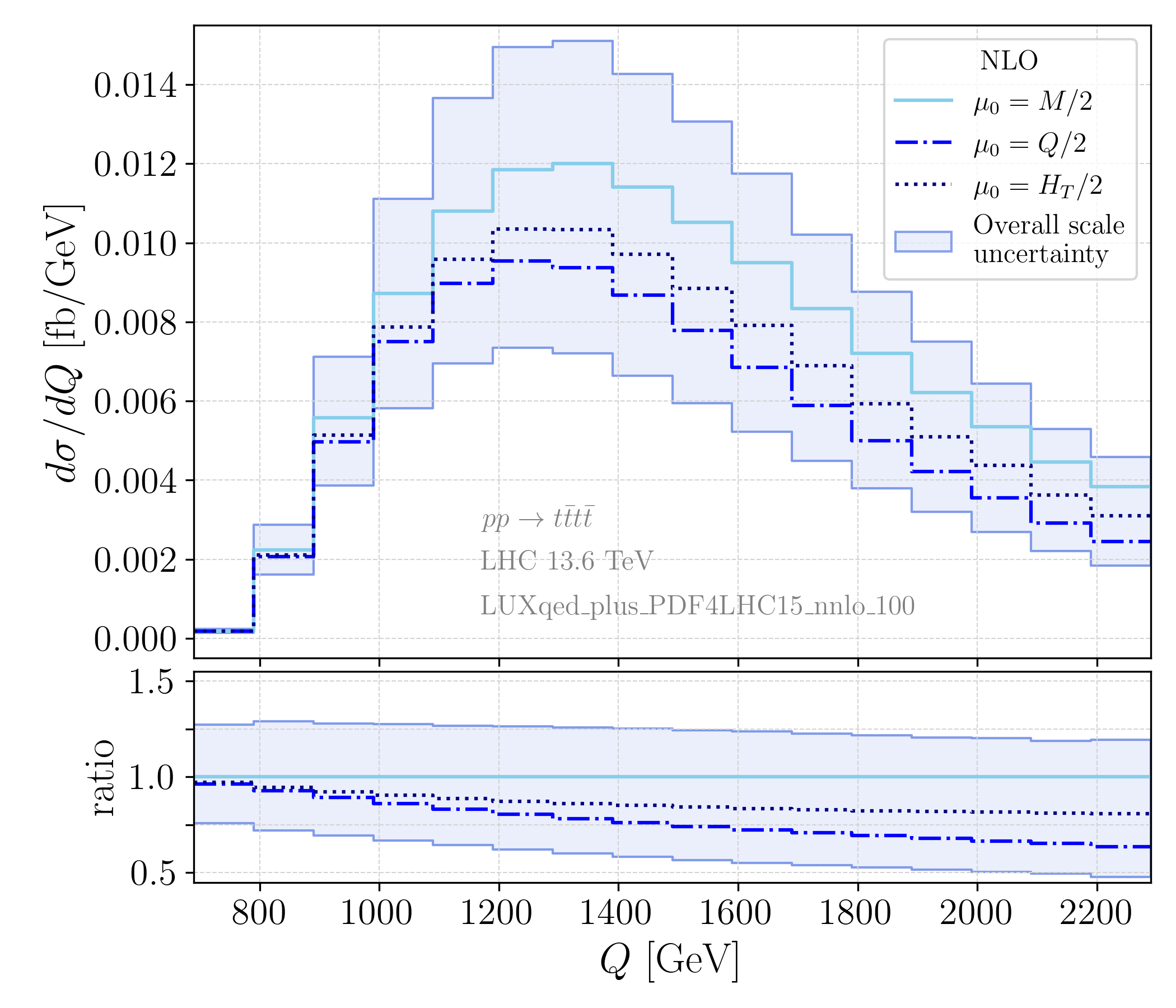}
    \includegraphics[width=0.45\textwidth]{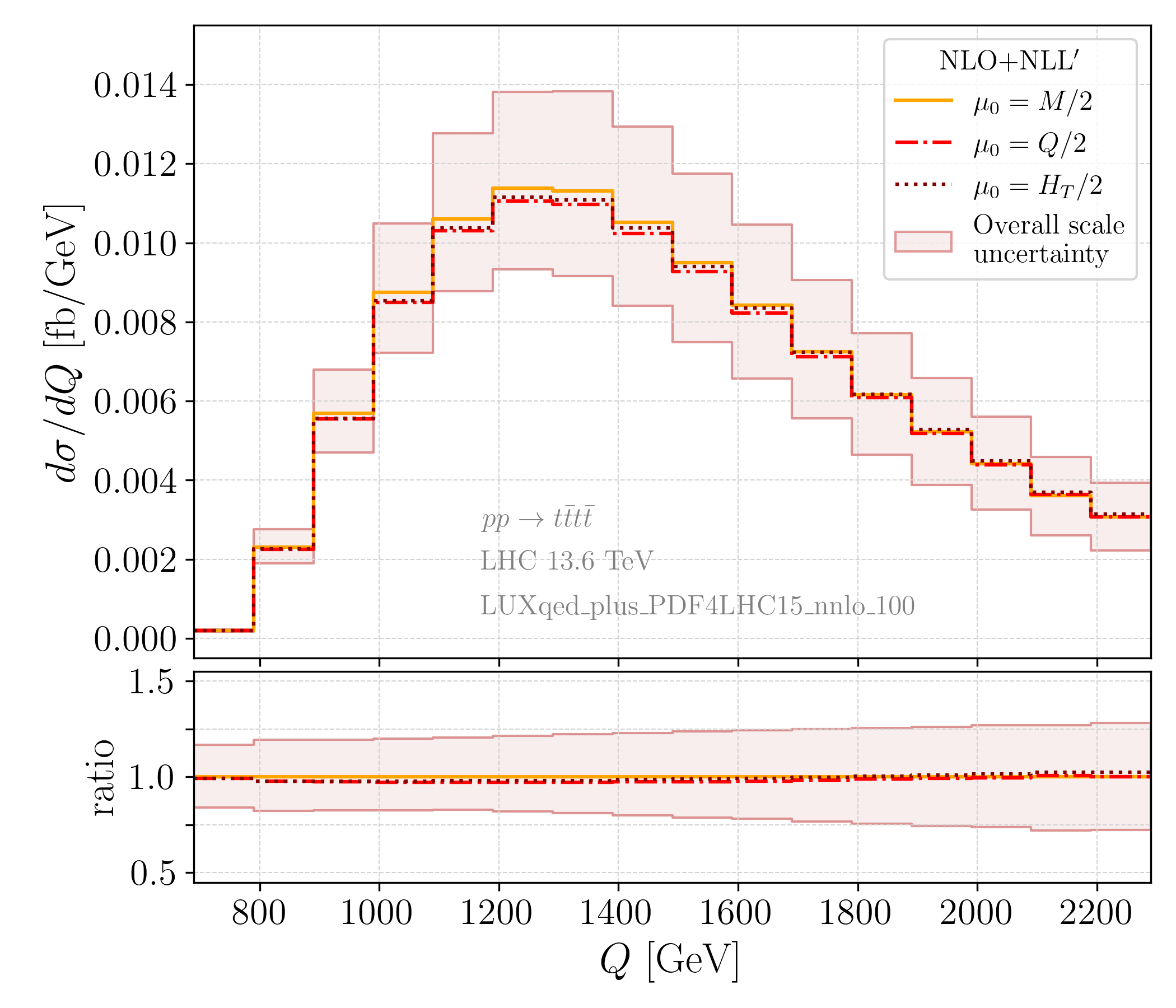}
    \caption{\it Invariant-mass distribution of the $t\bar{t}t\bar{t}$ system at NLO (left) and at NLO+NLL$'$ (right) accuracy.
    Distributions for three different scale choices are reported in each plot.
    The upper panels show the absolute distributions, the lower panels the distributions normalised bin-by-bin to the $\mu_0 = M/2$ distribution.
    The envelopes reflect the overall uncertainty by accounting for variations across all three scales considered.
    }
    \label{fig:DiffDistr_accuracies}
\end{figure}

This can be seen by grouping together the NLO (left) and NLO+NLL$'$ (right) results, as shown in Fig.~\ref{fig:DiffDistr_accuracies}.
As before, the top panel displays the invariant-mass distribution, but the lower panel now shows the ratio to the result obtained with $\mu_0 = M/2$. 
Secondly, in contrast to before, the shaded envelope now represents the combined scale uncertainty across all three scales, with its upper and lower bounds defined by the largest and smallest values obtained from the individual 7-point variations.
%
These plots clearly show the improvements introduced by the NLL$'$ corrections, as the predictions display a better convergence and a lower overall scale uncertainty.
Indeed, as illustrated in the lower panels of Fig.~\ref{fig:DiffDistr_accuracies}, the NLO predictions differ between the values calculated with different central scale choices up to $36\%$ for the central values, while they are $3\%$ at most once the NLL$'$ soft-gluon corrections are included.
Similar conclusions apply to the distributions obtained by matching to the NLO QCD without EW corrections.

\subsection{Total cross section}
\label{sec:total}
\begin{figure}[t]
    \centering
    \includegraphics[width=\textwidth]{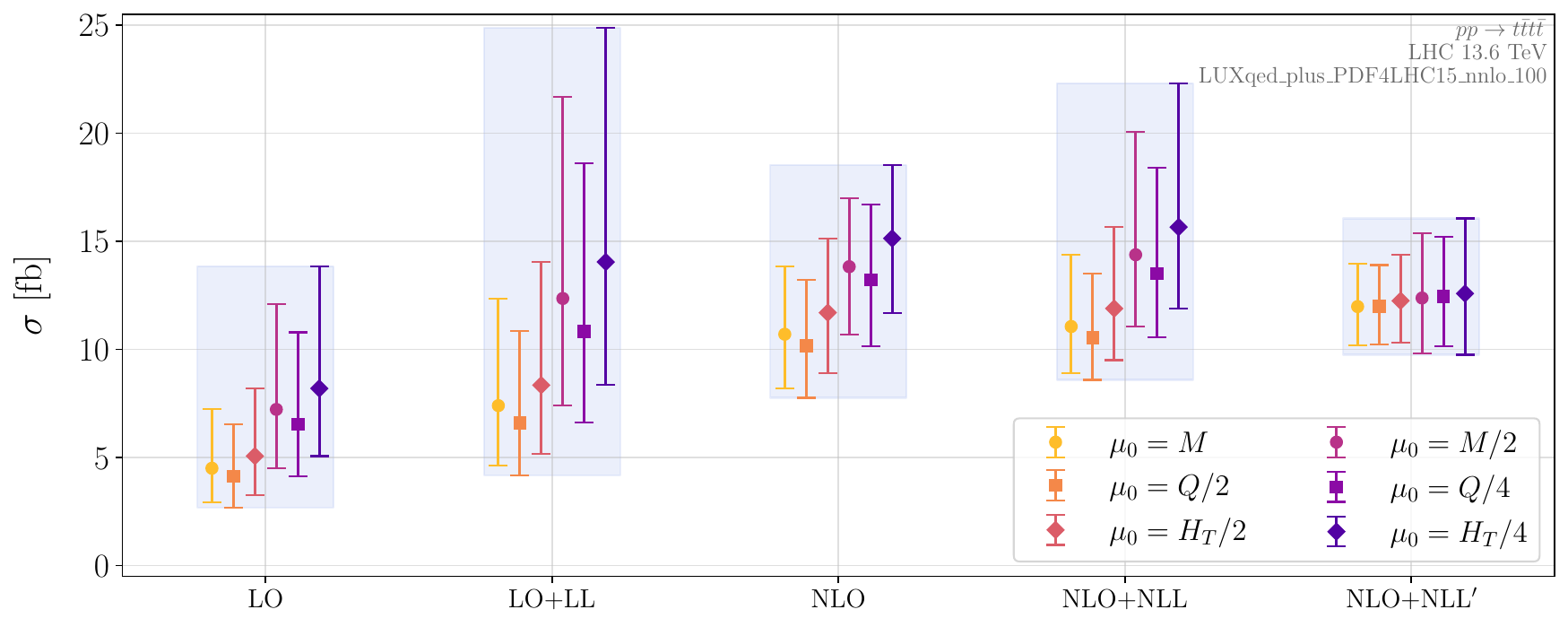}
    \caption{\it Predictions for the total cross section for the $pp\rightarrow t\bar{t}t\bar{t}$ process for fixed-order calculations and resummation-improved results, using different scale choices.
    The different-coloured error bars show the scale uncertainty for each scale choice, whereas the blue shaded region reflects the overall uncertainty by accounting for variations across all the considered scale choices.}
    \label{fig:stability_xs}
\end{figure}

We now turn to the results for the total cross section, obtained by integrating over the invariant-mass distributions.
Fig.~\ref{fig:stability_xs} shows the total cross section for five different levels of accuracy: LO, LO+LL, NLO, NLO+NLL and NLO+NLL$'$. 
The six different colours represent six different values for the central scale choice, where we take as before $\mu_0 = M/2$ (pink), $\mu_0=Q/2$ (orange) and $\mu_0 = H_T/2$ (red), but also add $\mu_0 = M$ (yellow), $\mu_0 = Q/4$ (violet), and $\mu_0 = H_T/4$ (purple), with corresponding error bars indicating the 7-point scale uncertainty obtained with only that central scale choice.
The blue shaded region corresponds to the maximum and minimum values for the cross sections across all considered scale variations. 
Similarly to the invariant-mass distribution, also for the total cross section we note that the inclusion of NLL$'$ corrections leads to a reduction in the scale dependence of the individual predictions.
We also observe good convergence across all six central scale choices, resulting in a narrower blue shaded region.\footnote{We verified that this convergence is driven by the term proportional to the virtual corrections $\mathbf{V}^{(1)}$ in the resummation formulae, with the latter included in the definition of $\mathbf{H}^{(1)}$, see Eq.~\eqref{eq:H1eqV1plusC1}.
The fact that the NLO+NLL$'$ predictions are less dependent on $\mu_R$ and $\mu_F$ has already been noted for associated $t\bar{t}$ production, see e.g. \cite{Kulesza:2017ukk,Kulesza:2018tqz,Kulesza:2020nfh}.}
%

%

We conclude this section by reporting in Tab.~\ref{tab:intXS} the numerical values for the total cross sections, some of which correspond to the points in Fig.~\ref{fig:stability_xs}.
Specifically, we show the NLO, NLO+NLL and NLO+NLL$'$ results for three central scale choices ($\mu_0 = Q/2,\, M/2,\, H_T/2$), both for $\sqrt{S} = 13.6$ and $13$ TeV.
We also report the $\mathcal{K}$-factor for each prediction w.r.t.~the corresponding NLO result.
The results presented in the top block of Tab.~\ref{tab:intXS} are obtained by matching to the full NLO predictions.
The NLL$'$ corrections are $+18\%$ for $\mu_0 = Q/2$, $-10\%$ for $\mu_0 = M/2$, and $+5\%$ for $\mu_0 = H_T/2$ at $13.6$ TeV.
We note that the NLL$'$ corrections obtained in the invariant-mass threshold formalism for the scale $\mu_0 = M/2$ have opposite sign compared to those obtained in the absolute-mass threshold formalism used in Ref.~\cite{vanBeekveld:2022hty}.
This reference reported a NLL$'$ correction of the size close to the one now observed for $\mu_0 = Q/2$.
We comment on this further in Appendix~\ref{sec:InvvsAbs}.
The bottom block of Tab.~\ref{tab:intXS} displays results obtained by matching to pure NLO QCD cross sections (NLO\textsubscript{QCD}), highlighting the effect of soft-gluon corrections.
These results constitute the most precise QCD predictions for four top-quark production currently available.
By comparing the two blocks, we observe that the EW corrections provide an overall increase of the total cross section values by $5$-$8\%$.

%
\renewcommand{\arraystretch}{1.25}
\begin{table}[t]
	\centering
	\begin{tabular}{ccccccc}
		\hline\hline
		$\sqrt{S}$ [TeV] & $\mu_0$ & NLO [fb] & NLO+NLL [fb] & $\mathcal{K}^{\text{NLL}}$ & NLO+NLL$'$ [fb] & $\mathcal{K}^{\text{NLL}'}$ \\
		\hline\hline
		$13.6$ & $M/2$ & $13.83_{-22.6\%}^{+23.0\%}$ & $14.38_{-23.1\%}^{+39.6\%}$ & 1.04 & $12.38_{-20.8\%}^{+24.1\%}$ & 0.90 \\
		& $Q/2$ & $10.16_{-23.6\%}^{+30.1\%}$ & $10.55_{-18.6\%}^{+27.9\%}$ & 1.04 & $12.00_{-14.9\%}^{+15.8\%}$ & 1.18 \\
		& $H_T/2$ & $11.70_{-23.8\%}^{+29.3\%}$ & $11.89_{-20.1\%}^{+31.7\%}$ & 1.02 & $12.25_{-15.7\%}^{+17.3\%}$ & 1.05 \\
        \hline
		$13$ & $M/2$ & $11.61_{-22.7\%}^{+22.9\%}$ & $12.05_{-23.0\%}^{+39.4\%}$ & 1.04 & $10.43_{-20.8\%}^{+23.6\%}$ & 0.90 \\
		& $Q/2$ & $8.56_{-23.7\%}^{+30.2\%}$ & $8.90_{-18.6\%}^{+27.9\%}$ & 1.04 & $10.16_{-14.8\%}^{+15.7\%}$ & 1.19 \\
		& $H_T/2$ & $9.84_{-23.9\%}^{+29.4\%}$ & $10.00_{-20.0\%}^{+31.5\%}$ & 1.02 & $10.35_{-15.7\%}^{+17.1\%}$ & 1.05 \\
		\hline\hline
		$\sqrt{S}$ [TeV] & $\mu_0$ & NLO\textsubscript{QCD} [fb] & NLO\textsubscript{QCD}+NLL [fb] & $\mathcal{K}^{\text{NLL}}$ & NLO\textsubscript{QCD}+NLL$'$ [fb] & $\mathcal{K}^{\text{NLL}'}$ \\
		\hline\hline
		$13.6$ & $M/2$ & $13.13_{-24.5\%}^{+25.2\%}$ & $13.68_{-24.9\%}^{+42.6\%}$ & 1.04 & $11.68_{-21.7\%}^{+26.2\%}$ & 0.89 \\
		& $Q/2$ & $9.38_{-25.8\%}^{+33.3\%}$ & $9.77_{-20.2\%}^{+30.7\%}$ & 1.04 & $11.22_{-16.2\%}^{+17.3\%}$ & 1.20 \\
		& $H_T/2$ & $10.88_{-25.8\%}^{+32.3\%}$ & $11.08_{-21.8\%}^{+34.7\%}$ & 1.02 & $11.43_{-16.8\%}^{+19.0\%}$ & 1.05 \\
        \hline
		$13$ & $M/2$ & $11.00_{-24.4\%}^{+24.9\%}$ & $11.44_{-24.7\%}^{+42.1\%}$ & 1.04 & $9.82_{-21.8\%}^{+25.5\%}$ & 0.89 \\
		& $Q/2$ & $7.90_{-25.8\%}^{+33.3\%}$ & $8.24_{-20.2\%}^{+30.6\%}$ & 1.04 & $9.50_{-16.1\%}^{+17.1\%}$ & 1.20 \\
		& $H_T/2$ & $9.14_{-25.9\%}^{+32.1\%}$ & $9.31_{-21.7\%}^{+34.4\%}$ & 1.02 & $9.66_{-16.8\%}^{+18.6\%}$ & 1.06 \\
		\hline
	\end{tabular}
	\caption{\it Theoretical predictions for the integrated cross section of $pp\rightarrow t\bar{t}t\bar{t}$ at several levels of accuracy and for different scale choices, both for 13 and 13.6 TeV LHC.
		The theoretical uncertainty from scale variation and the size of the NLL and NLL$'$ soft-gluon corrections are reported.}
	\label{tab:intXS}
\end{table}

%
%
\section{Summary}
\label{sec:sum}
In this study, we applied for the first time the invariant-mass threshold resummation formalism to a process with six coloured particles.
We computed the total cross section and invariant-mass distribution for the production of four top quarks at the LHC at NLO+NLL$'$ accuracy. 
Predictions for LHC centre-of-mass energy of $13.6$~TeV for different central scale choices have been obtained. For the total cross section we provided predictions also for $13$~TeV.
We also discussed the treatment of Coulomb singularities appearing in the soft anomalous dimension when a subset of final-state particles are produced at rest. 

We first presented results for the invariant-mass distribution of the $t\bar{t}t\bar{t}$ system.
The inclusion of NLL$'$ corrections reduces the theoretical uncertainty for the three scales considered  ($M/2$, $Q/2$ and $H_T/2$).
Moreover, they introduce relevant shape modifications that highly depend on the central scale employed.
Additionally, we have observed a significant improvement in the convergence  of the predictions obtained with  the three central scales.
The maximum spread among the predictions decreases from $36\%$ at NLO to just $3\%$ when including NLL$'$ corrections in the range of invariant masses considered.
Similar conclusions apply to the integrated cross section.
The convergence is in fact observed for a wider range of scales, making the scale choice at NLO+NLL$'$ much less relevant than that at NLO.

To conclude, resumming soft-gluon corrections at NLL$'$ accuracy, together with matching to the NLO calculation, leads to improved predictions for the cross section and invariant-mass distribution.
Specifically, we obtained the most accurate QCD predictions for $\ftop$ production to date.
%
%
\acknowledgments
%
We thank Federico Buccioni for providing the  colour-decomposed one-loop amplitudes in the \textsf{OpenLoops} software package. MvB is supported by the Dutch Research Council (NWO) under  project number VI.Veni.232.190. 
Part of the calculations for this publication were performed on the HPC cluster PALMA II of the University of Münster, subsidised by the DFG (INST 211/667-1).

\appendix

\section{Comparison between absolute-mass and invariant-mass threshold resummation}
\label{sec:InvvsAbs}
The absolute-mass threshold resummation (AMT-res) and invariant-mass threshold resummation (IMT-res) are two approaches to take into account soft-gluon corrections to all orders in QCD.
They are compared in this appendix.

In AMT-res, one computes the soft-gluon corrections originating from the phase-space region close to the four-top \emph{production threshold} $M = 4 m_t$, whereas IMT-res regards soft-gluon corrections from the phase-space region close to the four-top invariant mass $Q$.
Since AMT-res applies to total cross sections, dynamical scales such as $Q$ or $H_T$ are not adequate renormalisation or factorisation scale choices for AMT-res calculations, leaving scales proportional to $M$ as possible options in this approach. Following the choice made in the AMT-res calculations in  Ref.~\cite{vanBeekveld:2022hty}, we perform the comparison of the two approaches using $\mu_0 = M/2$.
In Tab.~\ref{tab:AMT-res_IMT-res_nll} we report cross sections values for different resummation accuracies in the two approaches. 
We also consider a specific (fictitious) modification to the IMT-res approach, labeled IMT-res\textsuperscript{test}, whose significance will be explained in what follows. 
\renewcommand{\arraystretch}{1.25}
\begin{table}[t]
	\centering
	\begin{tabular}{cccccc}
		\hline
		\hline
		\small{NLO} & \multicolumn{2}{c}{\small{$\sqrt{S} = 13.6$ TeV$\quad-\quad \mu_0 = M/2$ }} & \multicolumn{3}{c}{\small{$-\qquad$ Cross sections  in [fb]}} \\
		\hline
		$13.83_{-22.6\%}^{+23.0\%}$ &  &  &  &  & \\
		\hline
		\hline
		\small{NLO+NLL\textsubscript{IMT-res}} & \small{$\mathcal{K}$\textsubscript{IMT-res}} & \small{$\text{NLO+NLL}^\text{test}_\text{IMT-res}$} & \small{$\mathcal{K}^\text{test}_\text{IMT-res}$} & \small{NLO+NLL\textsubscript{AMT-res}} & \small{$\mathcal{K}$\textsubscript{AMT-res}} \\
		\hline
		$14.38_{-23.1\%}^{+39.6\%}$ & $1.04$ & $14.75_{-16.7\%}^{+19.7\%}$ & $1.07$ & $14.69_{-16.4\%}^{+16.8\%}$ & $1.06$ \\
		\hline
		\hline
		\small{NLO+NLL$'$\textsubscript{IMT-res}} & \small{$\mathcal{K}$\textsubscript{IMT-res}} & \small{$\text{NLO+NLL}$$'$$^{\,\text{test}}_\text{IMT-res}$} & \small{$\mathcal{K}^\text{test}_\text{IMT-res}$} & \small{NLO+NLL$'$\textsubscript{AMT-res}} & \small{$\mathcal{K}$\textsubscript{AMT-res}} \\
		\hline
		$12.38_{-20.8\%}^{+24.1\%}$ & $0.90$ & $17.91_{-20.8\%}^{+13.8\%}$ & $1.30$ & $17.36_{-17.5\%}^{+8.3\%}$  & $1.26$ \\
		\hline
		\hline
	\end{tabular}
	\caption{\it Total cross section values for $pp\rightarrow t\bar{t}t\bar{t}$.
    The theoretical uncertainties from scale variation are reported.
    The soft-gluon corrections are obtained from the IMT-res and AMT-res formalisms, as well as from a fictitious resummation IMT-res\textsuperscript{test}, whose definition is explained in the text.
	}
	\label{tab:AMT-res_IMT-res_nll}
\end{table}

One immediately observes that the corrections for AMT-res are systematically larger than those for IMT-res.
The former are $+6\%$ at NLL and $+26\%$ at NLL$'$, while the latter are $+4\%$ at NLL and become negative ($-10\%$) at NLL$'$.
The dominant part of the numerical differences between these results can be traced back to scales-ratio logarithms, involving the scales characterising the threshold,  in the jet collinear functions $\Delta_i$, cf. Eq.~\eqref{eq:collinearfunctions}. 
The scales-ratio logarithms in question appear starting from the NLL accuracy. In particular, in IMT-res the NLL $g_2(\lambda)$ function has the form\footnote{We do not report the expressions for $A^{(1)}$, $A^{(2)}$, $b_0$ and $b_1$, as they are not relevant for the purposes of the discussion at hand. They can be found, for instance, in Ref.~\cite{Catani:2003zt, vanBeekveld:2019cks}.}
\begin{align}
\nonumber
g_2(\lambda) & = \frac{A^{(1)} b_1}{2 \pi b_0^3}\left[ 2\lambda + \log(1-2\lambda) + \frac{1}{2}\log^2(1-2\lambda) \right]
       - \frac{A^{(2)}}{2 \pi^2 b_0^2}\left[ 2\lambda + \log(1-2\lambda) \right]\\
       & + \frac{A^{(1)}}{2 \pi b_0} \left[ \log(1-2\lambda) \log \left(\frac{Q^2}{\mu_R^2}\right) + 2 \lambda \log\left(\frac{\mu_F^2}{\mu_R^2}\right) \right].
       \label{eq:g2}
\end{align}
In AMT-res, $\log Q^2/\mu_{R}^2$ is replaced by $\log M^2/\mu_{R}^2$. It is the difference between the values of these logarithmic terms which drives the numerical differences between the results in the two approaches. After all, the invariant mass of the four-top system is equal to the absolute-mass threshold value only in the strict limit of production of four top quarks at rest and, as shown in Fig.~\ref{fig:DiffDistrHT2}, the bulk of the contributions to the total cross section comes from the region of substantially higher $Q$'s.
To explicitly see this effect, we introduce the fictitious modification IMT-res\textsuperscript{test} where we replace $Q^2$ with $M^2$ in the $g_2(\lambda)$ function used in IMT-res, Eq.~\ref{eq:g2}, (and the corresponding term appearing in the NLO expansion), but leave all other ingredients untouched. 
Indeed, based on Tab.~\ref{tab:AMT-res_IMT-res_nll} we observe that the $\mathcal{K}$-factors substantially increase after this replacement, getting close to the AMT-res ones, both at NLO+NLL and especially at NLO+NLL$'$ accuracy, where the original gap between the results was larger.
Other ingredients that differ between IMT-res and AMT-res, such as the soft-anomalous dimension matrix, are found to lead to numerically less significant differences.
We conclude the discussion of the $\mathcal{K}$-factors by noticing that the NLL$'$ AMT-res corrections ($+26\%$) are similar in size to the NLL$'$ IMT-res corrections obtained with the scale choice $\mu_0 = Q/2$ ($+18\%$, shown in Tab.~\ref{tab:intXS}). As each result was computed with a $\mu_0$ scale proportional to the the scale characterising the respective production threshold, the similarity in the size of corrections is a manifestation of theoretical consistency between the two approaches.

To see that the difference between AMT-res and IMT-res indeed grows as the value of $Q$ departs from $M$, we turn now to the $\ftop$ invariant-mass distribution.
To this end, we first define the resummation correction that features in the additive matching procedure, similarly to Eq.~\eqref{eq:matching}, as 
 \begin{equation}
 	\label{eq:diff_matching}
 	\frac{\diff\sigma^{\text{res}-\text{exp}}}{\diff Q} \equiv \frac{\diff\sigma^{\text{res}}}{\diff Q} - \frac{\diff\sigma^{\text{res}}}{\diff Q}\bigg|_{\text{NLO}}\,,
 \end{equation}
where ``res$-$exp" stands for ``resummed minus expanded".
Strictly speaking, the invariant-mass distribution can only be calculated in the IMT-res formalism. However, we can construct an additional invariant-mass distribution by reweigthing the $Q$-dependent hard function with the AMT-res exponential factors. By construction,  when integrated over the entire $Q$ range, it will return the corresponding  AMT-res total cross section, in analogy to the IMT-res distribution. We stress though that this distribution, labelled ``AMT-res", is purely technical and only intended to attain better understanding of the differences between the two approaches.
For comparison we also consider the IMT-res\textsuperscript{test} distribution, obtained from the IMT-res distribution by modifying the logarithmic term in $g_2$, in analogy to the total cross section.

\begin{figure}[t]
	\centering
    \includegraphics[width=0.6\textwidth]{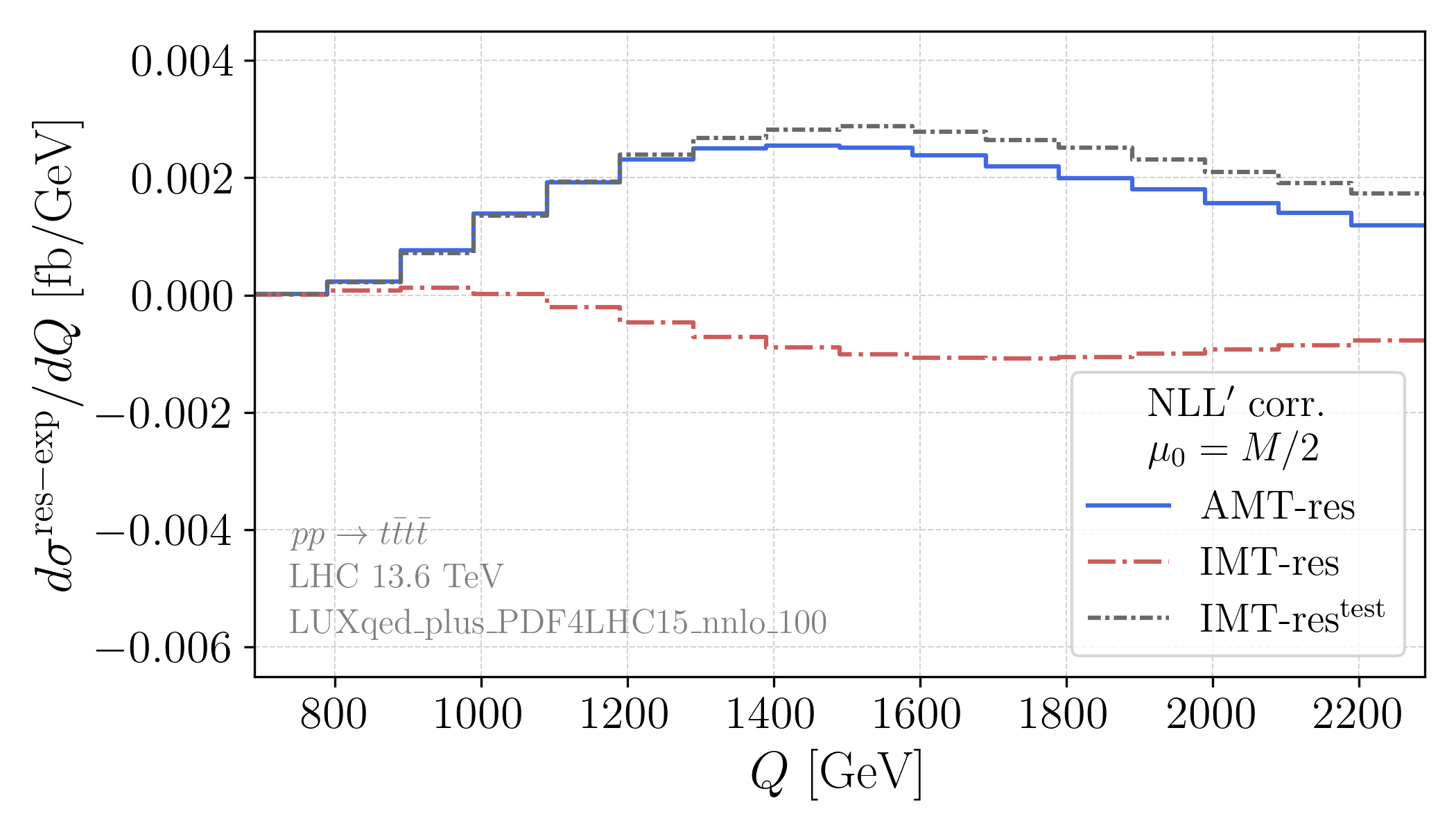}
	\caption{\it NLL$'$ corrections to the NLO invariant-mass distribution of the $t\bar{t}t\bar{t}$ system, as defined in Eq.~\eqref{eq:diff_matching}, in the AMT-res and IMT-res formalisms.
    The distribution for the fictitious resummation IMT-res\textsuperscript{test} (whose definition is explained in the text) is also provided.}
	\label{fig:AMT_IMT_diff_distr}
\end{figure}

In Fig.~\ref{fig:AMT_IMT_diff_distr}, we compare the correction \eqref{eq:diff_matching} for the IMT-res and the newly constructed AMT-res and IMT-res\textsuperscript{test} differential distributions at NLL$'$ accuracy and for $\mu_0 = M/2$.
One clearly sees that, for $Q \to M$, the IMT-res result approaches the AMT-res result. 
However, away from the absolute-mass threshold limit, the numerical difference between the two approaches grows and remains sizable, explaining the observed difference between the total cross sections. In particular, the IMT-res NLL$'$ correction obtained for $\mu_0=M/2$ becomes negative already at relative low values of $Q$ which leads to a $\mathcal{K}$-factor lower than $1$ for the corresponding total cross section.
Finally, the IMT-res\textsuperscript{test} distribution shows a close resemblance to the AMT-res one, in line with the behavior previously observed at the integrated level.

\captionsetup[subfigure]{labelformat=empty}  

\section{\texorpdfstring{Comparison between $\bar N$- and $N$- resummation}{Comparison between Nbar and N resummation}}
\label{sec:NbarvsN}
Throughout this work, resummation has been carried out in Mellin space using logarithms of $\bar{N}\equiv N e^{\gamma_E}$ rather than the Mellin moment $N$.
We refer to the former approach as ``$\bar{N}$-resummation" and to the latter as ``$N$-resummation".
In this Appendix, we motivate this choice and assess its impact in $\ftop$ production.

As mentioned in Sec.~\ref{sec:theory}, invariant-mass threshold resummation resums terms involving plus-distributions, which transform in Mellin space as
\begin{equation}
\label{eq:mellinmoment}
    I_m(N)\equiv\int_0^1 \diff \hat{\rho} \,\hat{\rho}^{N-1}\biggl[\frac{\log^m(1-\hat{\rho})}{1-\hat{\rho}}\biggr]_+ \,.
\end{equation}
The general solutions to such integrals are harmonic sums \cite{Vermaseren:1998uu,Moch:2005ba}.
In Mellin space, the threshold limit $\hat{\rho} \rightarrow 1$ corresponds to  $N\rightarrow \infty$, and Eq.~\eqref{eq:mellinmoment} expands as \cite{Catani:1989ne,Catani:2003zt}
\begin{equation}
\label{eq:logN_polynomial}
    I_m(N) \simeq \sum_{k=0}^{m+1} a_{km} \,(\log N + \gamma_E)^k + \mathcal{O}\biggl(\frac{1}{N}\biggr),
\end{equation}
where the $a_{km}$ coefficients are linear combinations of powers of Riemann zeta values \cite{Catani:1989ne}. Due to the presence of a non-logarithmic $\gamma_E$ term in Eq.~\eqref{eq:logN_polynomial}, when the binomial expansion is carried out for each $k$, factors with lower powers of $\log N$ are produced. Such terms are formally of higher logarithmic order compared to the $\log^k N$ term.
The form of Eq.~\eqref{eq:logN_polynomial} leads to the possibility of reorganising the resummation in terms of $\bar{N}$.

\begin{figure}[t]
    \centering
    \includegraphics[width=0.6\textwidth]{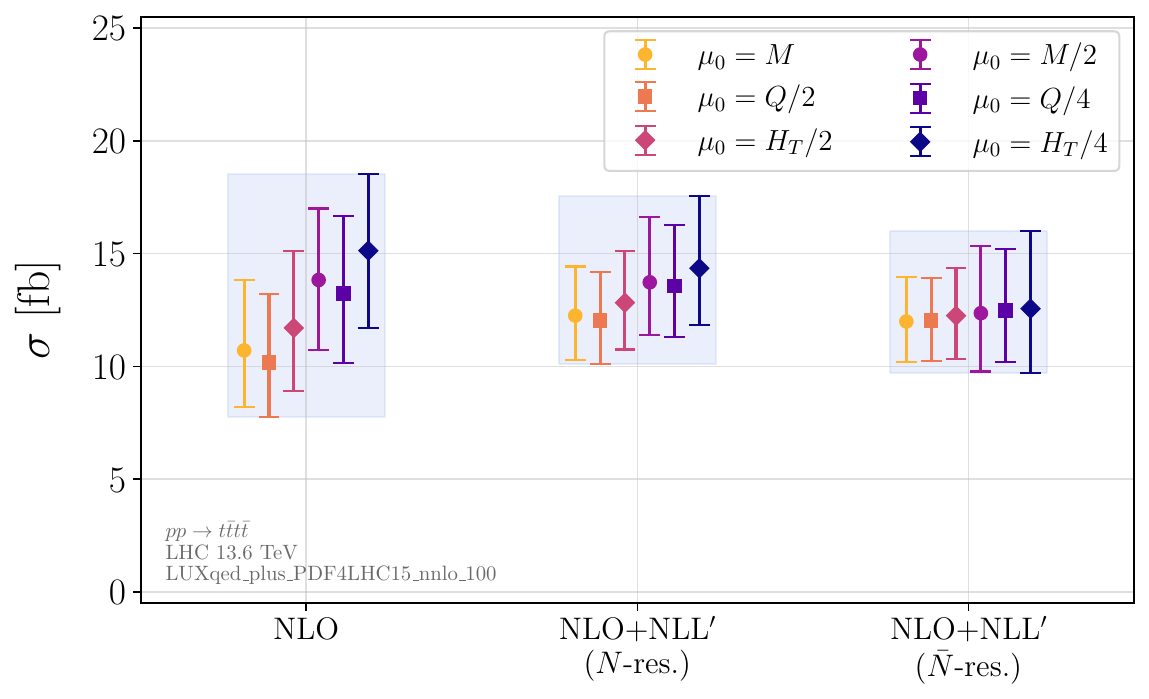}
    \caption{\it Predictions for the integrated cross section of $pp\rightarrow t\bar{t}t\bar{t}$ for different scale choices.
    The first panel reports the NLO predictions, the second panel NLO+NLL$'$ $N$-resummed predictions and the third panel NLO+NLL$'$ $\bar{N}$-resummed predictions.
    The error bars represent an estimate of the theoretical uncertainty, obtained through a $7$-point scale variation.
    The blue envelope reflects the overall uncertainty by accounting for variations across all three scales considered.}
    \label{fig:stability_xs_nbar-res}
\end{figure}

In the context of our study, working with $\bar{N}$-resummation allows us to capture some NNLL terms already at NLL$(')$ in the soft and jet functions.
These higher-order logarithmic contributions introduce small numerical differences between the two approaches in the predictions for the cross section.
In Fig.~\ref{fig:stability_xs_nbar-res} we compare the $\ftop$ production cross section at NLO+NLL$'$ in the two resummation approaches.
The predictions are compatible within uncertainties, but converge significantly better in the $\bar{N}$-resummation approach compared to the $N$-resummation case.
This behaviour was already observed for Drell-Yan production~\cite{AH:2020cok} and deep inelastic scattering~\cite{Das:2019btv}.\footnote{The differences between $N$-resummation and $\bar{N}$-resummation were also studied for Higgs-pair production at the LHC~\cite{AH:2022elh}.}
A similar trend is seen (but not reported here) also for the invariant-mass distribution.
Finally, we remark that the expansion of the NLL$'$ cross section up to NLO is identical in the two formalisms
\begin{equation}
    \tilde\sigma^{\rm N-res}|_{\rm NLO}=\tilde\sigma^{\rm \bar{N}-res}|_{\rm NLO}\,.
\end{equation}

%

\section{Approximate NLO results}
\label{sec:expanded}
\begin{figure}[t]
    \centering
    \includegraphics[width=0.6\textwidth]{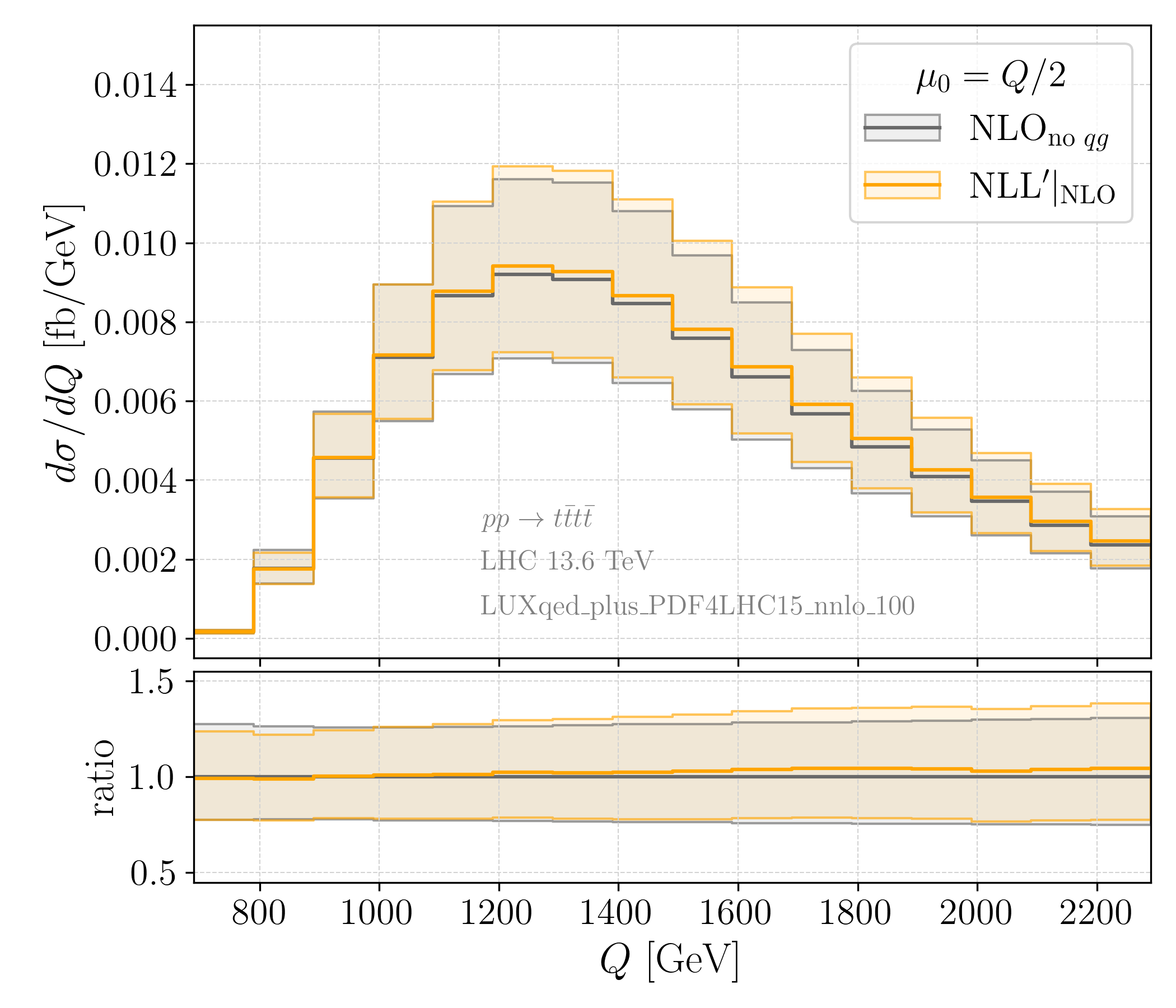}
    \caption{\it NLO\textsubscript{no $qg$} and NLL$'|$\textsubscript{NLO} invariant-mass distributions of the $t\bar{t}t\bar{t}$ system.
    The envelopes represent the uncertainty obtained by scale variation.
    The upper panel shows the absolute distributions, the lower panel the distributions normalised bin-by-bin to the NLO\textsubscript{no $qg$} distribution.
   }
    \label{fig:DiffDistr_NLO_v_Exp}
\end{figure}

In this Appendix we show that the expansion of the NLL$'$ resummed result up to NLO in QCD (NLL$'|$\textsubscript{NLO}) provides a sufficiently accurate approximation of the exact NLO\textsubscript{QCD} invariant-mass distribution.
Since in our formalism we do not resum the (anti-)quark-gluon initiated ($qg$) channels, as it is a next-to-leading power contribution, the appropriate baseline for this comparison is the NLO\textsubscript{QCD} result excluding these channels (NLO\textsubscript{no $qg$}).
Fig.~\ref{fig:DiffDistr_NLO_v_Exp} shows the comparison between the NLL$'|$\textsubscript{NLO} prediction (yellow) and the NLO\textsubscript{no $qg$} one (black).
%
%
Theoretical uncertainties from scale variation are reported along the central values of the distributions.
Results are provided for $\mu_0 = Q/2$, but similar conclusions hold for the other scale choices.
Once again, the upper panel presents the absolute value of the distributions, while the lower panel reports the ratio plots normalised to NLO\textsubscript{no $qg$}.
The latter highlights bin-by-bin the relative differences, which are small and vary from $-1\%$ to $4\%$.
Moreover, the uncertainty bands for NLL$'|$\textsubscript{NLO} and NLO\textsubscript{no $qg$} largely overlap, demonstrating that the NLL$'$ expansion reproduces NLO\textsubscript{no $qg$} reliably.
We verified that the $qg$ contribution to the cross section is very small, with relative differences between NLO\textsubscript{no $qg$} and NLO\textsubscript{QCD} reaching at most $8\%$ in the tail of the distribution.
These findings suggest that the NLL$'|$\textsubscript{NLO} provides a good approximation of the full NLO\textsubscript{QCD} distribution.

This observation carries over to the total cross section. 
In Tab.~\ref{tab:NLLpexp} we compare, for the three different scales and for $\sqrt{S} = 13.6$ TeV, predictions for the total cross section at NLL$'|$\textsubscript{NLO}, NLO\textsubscript{no~$qg$} and NLO\textsubscript{QCD}. 
%
\renewcommand{\arraystretch}{1.25}
\begin{table}[t]
\centering
\begin{tabular}{cccc}
\hline\hline
$\mu_0$ & NLO\textsubscript{QCD} [fb] & NLO\textsubscript{no~$qg$} [fb] &  NLL$'|$\textsubscript{NLO} [fb] \\
\hline\hline
$M/2$ & $13.13_{-24.5\%}^{+25.2\%}$ & $13.05_{-21.1\%}^{+20.2\%}$ & $13.45_{-21.9\%}^{+21.6\%}$ \\
$Q/2$ & $9.38_{-25.8\%}^{+33.3\%}$ & $9.77_{-23.9\%}^{+28.1\%}$ &
$9.92_{-24.1\%}^{+28.7\%}$ \\
$H_T/2$ & $10.88_{-25.8\%}^{+32.3\%}$ & $11.22_{-23.7\%}^{+26.0\%}$ &
$11.44_{-24.0\%}^{+27.0\%}$ \\
\hline
\end{tabular}
\caption{\it NLO, NLO excluding the (anti-)quark-gluon initiated channels and approximated NLO theoretical predictions for the integrated cross section of $pp\rightarrow t\bar{t}t\bar{t}$ at 13.6 TeV LHC for different scale choices. EW corrections are not included in this comparison.
The theoretical uncertainty from 7-point scale variation is reported.}
\label{tab:NLLpexp}
\end{table}
%
The differences between NLL$'|$\textsubscript{NLO} and NLO\textsubscript{no~$qg$} do not exceed $3\%$, while the differences between the approximated NLO and the exact NLO\textsubscript{QCD} are at most $6\%$, making NLL$'|$\textsubscript{NLO} a reliable approximation for the NLO\textsubscript{QCD} cross section.

%
\bibliography{References} 
%

\bibliographystyle{JHEP}

\end{document}